# Single-Atom Alloy Catalysts Designed by First-Principles Calculations and Artificial Intelligence

Zhong-Kang Han,[1,4] Debalaya Sarker,[1,4] Runhai Ouyang,[2,4] Yi Gao,[3]* Sergey V. Levchenko[1]*

[1] Center for Energy Science and Technology, Skolkovo Institute of Science and Technology, Skolkovo Innovation Center, Moscow, 143026, Russia
[2] Materials Genome Institute, Shanghai University, 333 Nanchen Road, Shanghai, 200444, P.R. China
[3] Shanghai Advanced Research Institute, Chinese Academy of Sciences, Shanghai, 201210, P. R. China
[4] These authors contributed equally.

Correspondence Email: S.Levchenko@skoltech.ru; gaoyi@zjlab.org.cn

**Abstract**

Single-atom metal alloy catalysts (SAACs) have recently become a very active new frontier in catalysis research. The simultaneous optimization of both facile dissociation of reactants and a balanced strength of intermediates' binding make them highly efficient and selective for many industrially important reactions. However, discovery of new SAACs is hindered by the lack of fast yet reliable prediction of the catalytic properties of the sheer number of candidate materials. In this work, we address this problem by applying a compressed-sensing data-analytics approach parameterized with density-functional inputs. Our approach is faster and more accurate than the current state-of-the-art linear relationships. Besides consistently predicting high efficiency of the experimentally studied Pd/Cu, Pt/Cu, Pd/Ag, Pt/Au, Pd/Au, Pt/Ni, Au/Ru, and Ni/Zn SAACs (the first metal is the dispersed component), we identify more than two hundred yet unreported candidates. Some of these new candidates are predicted to exhibit even higher stability and efficiency than the reported ones. Our study demonstrates the importance of breaking linear relationships to avoid bias in catalysis design, as well as provides a recipe for selecting best candidate materials from hundreds of thousands of transition-metal SAACs for various applications.

Recently, single-atom dispersion has been shown to dramatically reduce the usage of rare and expensive metals in heterogeneous catalysis, at the same time providing unique possibilities for tuning catalytic properties.[1, 2] The pioneering work by Sykes and co-workers[2] has demonstrated that highly dilute bimetallic alloys, where single atoms of Pt-group are dispersed on the surface of an inert metal host, are highly efficient and selective in numerous catalytic reactions. These alloy catalysts are now extensively used in the hydrogenation-related reactions such as hydrogenation of $CO_2$, water-gas shift reaction, hydrogen separation, and many others.[3-5] The outstanding performance of SAACs is attributed to a balance between efficiency of $H_2$ dissociation and binding of H at the surface of metallic alloys.[2, 6, 7]

Using desorption measurements in combination with high-resolution scanning tunneling microscopy, Kyriakou *et al*. have shown that isolated Pd atoms on a Cu surface can substantially reduce the energy barrier for both hydrogen uptake and subsequent desorption from the Cu metal surface.[2] Lucci and co-workers have observed that isolated Pt atoms on the Cu(111) surface exhibit stable activity and 100% selectivity for the hydrogenation of butadiene to butenes.[8] Liu *et al.* have investigated the fundamentals of CO adsorption on Pt/Cu SAAC using a variety of surface science and catalysis techniques. They have found that CO binds more weakly to single Pt atoms in Cu(111), compared to larger Pt ensembles or monometallic Pt. Their results demonstrate that SAACs offer a new approach to design CO-tolerant materials for industrial applications.[9] To date, Pd/Cu,[10-12] Pt/Cu,[7-9, 13-15] Pd/Ag,[12, 16] Pd/Au,[12] Pt/Au,[17] Pt/Ni,[18] Au/Ru,[19] and Ni/Zn[20] SAACs have been synthesized and found to be active and selective towards different hydrogenation reactions. However, the family of experimentally synthesized SAACs for hydrogenation remains small and comparisons of their catalytic properties are scarce.

Conventional approaches to design single-atom heterogeneous catalysts for different industrially relevant hydrogenation reactions mainly rely on trial-and-error methods. However, challenges in synthesis and *in situ* experimental characterization of SAACs impose limitations on these approaches. With advances in first-principles methods and computational resources, theoretical modeling is proven to provide new opportunities for rational catalyst design.[6, 21-48] The general simple yet powerful approach is the creation of a large database with first-principles based inputs, followed by intelligent interrogation of the database in search of materials with the desired properties.[35, 48] Significant efforts have been made in developing reliable descriptor-based models following the above general approach.[6, 21-35, 48] In catalysis, a descriptor is a parameter (a feature) of the catalytic material that is easy to evaluate and is correlated with a complex target property (e.g., activation energy or turnover frequency of a catalytic reaction). A notable amount of research has been devoted to near-linear dependencies between descriptors and target properties.[22-30] For example, the linear relationship between the reaction energies and the activation energies is known as the Brønsted-Evans-Polanyi relationship (BEP) in heterogeneous catalysis.[29, 30, 45-47] Also, the linear correlation between $d$-band center of a clean transition-metal surface and adsorption energies of molecules on that surface have been studied in great detail and widely applied.[22-24, 36, 44] In catalysis, near-linear correlations between adsorption energies of different adsorbates are referred to as scaling relations.[26, 28, 37] The advantages of such correlations are their simplicity and usually strong physical foundations. However, they are not exact, and there is an increasing number of studies focused on overcoming limitations imposed by the corresponding approximations.[6, 31-34, 38-41, 48] The nonlinear and intricate relationship between the catalysts' properties and surface reactions at realistic conditions[42, 43] has held back the reliable description of catalytic properties. Note that, although the stability of SAACs is of no less significance in designing a potential catalyst than their catalytic performance, it hasn't received the same attention.

In this work, combining first-principles calculations and compressed-sensing data-analytics methodology, we address the issues that inhibit the wider use of SAAC in different industrially important reactions. By identifying descriptors based only on properties of the host surfaces and guest single atoms, we predict the binding energies of H ($BE_H$), the dissociation energy barriers of $H_2$ molecule ($E_b$), and the segregation energies (SE) of the single guest atom at different transition metal surfaces. The state-of-the-art compressed-sensing based approach employed here for identifying the key descriptive parameters is the recently developed SISSO (sure independence screening and sparsifying operator).[49] SISSO enables us to identify the best low-dimensional descriptor in an immensity of offered candidates. The computational time required for our models to evaluate the catalytic properties of a SAAC is reduced by at least a factor of one thousand compared to first-principles calculations, which enables high-throughput screening of a huge number of SAAC systems.

**Results**

The $BE_H$ for more than three hundred SAACs are calculated within the framework of DFT with RPBE exchange-correlation functional. This large dataset consists of $BE_H$ values at different low-index surface facets including fcc(111), fcc(110), fcc(100), hcp(0001), and bcc(110) and three stepped surface facets including fcc(211), fcc(310), and bcc(210) of SAACs with twelve transition-metal hosts (Cu, Zn, Cr, Pd, Pt, Rh, Ru, Cd, Ag, Ti, Nb, and Ta). On each TM host surface, one of the surface atoms is substituted by a guest atom to construct the SAACs. H atom is placed at different non-equivalent high-symmetry sites close to the guest atom [Figure S1 in Supplementary Information (SI)], and the $BE_H$ for the most favorable site is included in the data set. Complete information on adsorption sites and the corresponding $BE_H$ is given in SI. The $BE_H$ are further validated by a comparison with previous calculations.[6, 21]

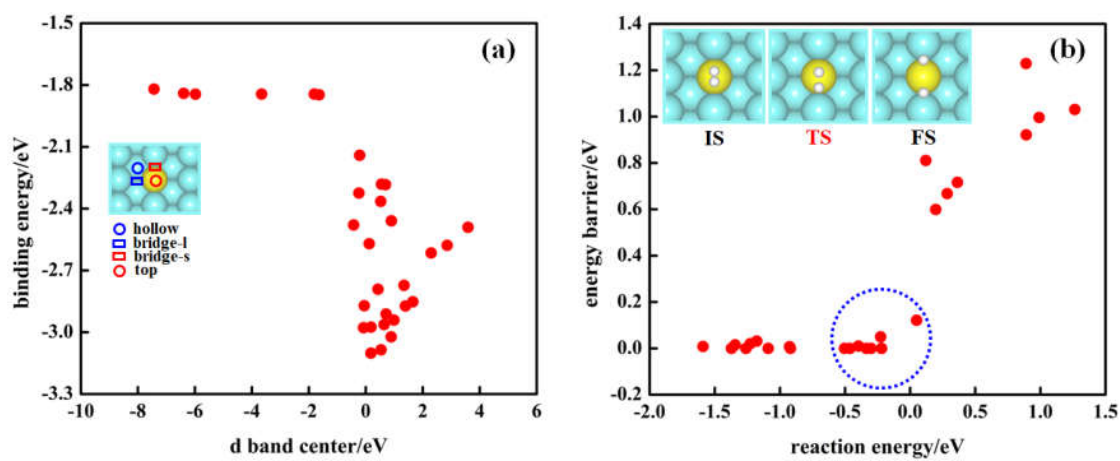

**Figure 1.** Correlation between (a) H-atom binding energy $BE_H$ and the $d$-band center and (b) the $H_2$ dissociation energy barrier $E_b$ and the $H_2$ dissociation reaction energy for Ag(110) based SAACs. The SAACs inside the blue dotted circle in (b) significantly reduce $E_b$ while reducing reaction energy only moderately.

To better understand the variation in $BE_H$ for different guest atoms, we first investigate correlation between $BE_H$ and the $d$-band center for alloyed systems. The correlation is shown in Figure 1a for different SAACs on Ag(110) host surface. According to the $d$-band center theory,[21, 23, 36, 44] the closer the $d$-band center is to the Fermi level, the stronger the $BE_H$ should be. However, it is evident from Figure 1a that the expected linear correlation, as predicted by the $d$-band model, is broken for SAACs for H adsorption. This is due to the small size of the atomic H orbitals, leading to a relatively weak coupling between H $s$ and the TM $d$-orbitals.[21] Furthermore, we check the validity of the BEP relations between the $E_b$ and the $H_2$ dissociation reaction energy for SAACs (Figure 1b), which is commonly used to extract kinetic data for a reaction on the basis of the adsorption energies of the reactants and products.[29, 45-47] As shown in Figure 1b, the highlighted SAACs inside the blue dotted circle significantly reduce $E_b$ while reducing reaction energy only moderately. As a result, SAACs provide small reaction energy and low activation energy barrier, which leads to breaking BEP relations and thus optimized catalytic

performance. The BEP relations are also found to be broken for other reactions catalyzed by SAACs.[6]

**Table 1.** Primary features used for the descriptor construction.

| system | class | name | abbreviation |
|---|---|---|---|
| host | atomic | Energy of the highest-occupied Kohn-Sham level | H* |
| | | Energy of the lowest-unoccupied Kohn-Sham level | L* |
| | | Electron affinity (Atomic radius) | EA*(R*)[#] |
| | | Ionization potential | IP* |
| | | Binding energy of H with single host metal atom (Binding energy of host metal dimers) | EH*(EB*)[#] |
| | | Binding distance of H with single host metal atom (Binding distance of host metal dimer) | dH*(dd*)[#] |
| | bulk | Cohesive energy | EC* |
| | | $d$-band center | DC* |
| | surface[¶] | $d$-band center of the top surface layer | DT* |
| | | $d$-band center of the subsurface layer | DS* |
| | | Slab Fermi level | F* |
| guest atom | atomic | Energy of the highest-occupied Kohn-Sham level | H |
| | | Energy of the lowest-unoccupied Kohn-Sham level | L |
| | | Electron affinity (Atomic radius) | EA(R)[#] |
| | | Ionization potential | IP |
| | | Binding energy of H with single guest metal atom (Binding energy of guest metal dimers) | EH(EB)[#] |
| | | Binding distance of H with single guest metal atom (Binding distance of guest metal dimers) | dH(dd)[#] |
| | bulk | Cohesive energy | EC |
| | | $d$-band center | DC |

[#]the feature in parentheses is used for the model of segregation energy (SE), while the feature outside parentheses is used for the models of H binding energy ($BE_H$) and $H_2$ dissociation energy barrier ($E_b$). [¶]the surface-based primary features were calculated using the slab unit cell consisting of one atom per atomic layer.

Thus, the standard simple correlations (from $d$-band center theory and the BEP relations) fail for H adsorption on SAACs. Moreover, the calculation of the $d$-band center for each SAAC is highly computationally demanding, considering the very large number of candidates. These facts emphasize the necessity to find new accurate but low-cost descriptors for computational screening of SAACs. In the SISSO method, a huge pool of more than ten billion candidate features is first constructed iteratively by combining 19 low-cost primary features listed in Table 1 using a set of mathematical operators. A compressed-sensing based procedure is used to select one or more most relevant candidate features and construct a linear model of the target property (see SI for details on the SISSO procedure). Note that the three primary surface features are properties of the pure host surfaces (elemental metal systems). This is undoubtedly much more efficient than obtaining the properties of SAACs (alloyed metal systems). In the latter case, due to the interaction between the single guest atom and its images, a large supercell of the whole periodic system containing guest atom and host surface needs to be computed. On the contrary, only smallest unit cell is needed to compute the pristine surface features.

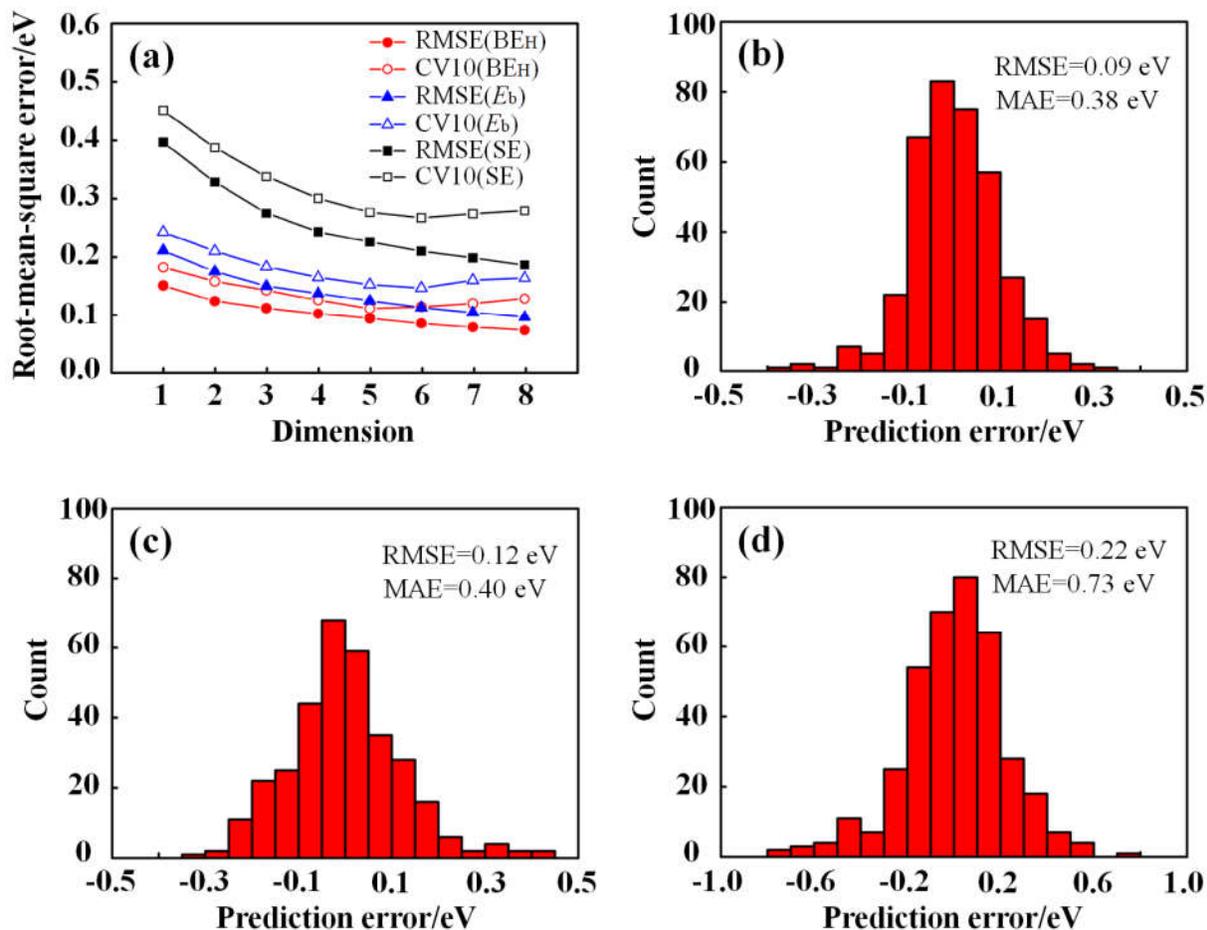

**Figure 2.** (a) RMSE and the averaged RMSE of the 10 fold cross-validation. (b-d) Distribution of errors for the best models versus RPBE results for $BE_H$ (b), $E_b$ (c), and SE (d).

To test the predictive power of obtained models, we employ 10-fold cross validation (CV10). The dataset is first split into 10 subsets, and the descriptor identification along with the model training is performed using 9 subsets. Then the error in predicting properties of the systems in the remaining subset is evaluated with the obtained model.[50-52] The CV10 error is defined as the average value of the test errors obtained for each of the ten subsets. In SISSO over-fitting may occur with increasing dimensionality of the descriptor (i.e., the number of complex features that are used in construction of the linear model).[49] The descriptor dimension at which the CV10 error starts increasing identifies the optimal dimensionality of the descriptor (the detailed validation approaches have been discussed in the SI to confirm its reliability). The root-mean-square errors (RMSE), together with the CV10 errors of the SISSO models for $BE_H$, $E_b$, and SE are displayed in Figure 2a. The obtained optimal descriptor dimensionalities for $BE_H$, $E_b$, and SE of the SAACs are 5, 6, and 6, respectively. Distribution of errors for the best models versus RPBE results is displayed in Figure 2(b-d). The RMSE, and maximum absolute error (MAE) of the models are also shown. The error distributions for all the lower-dimensional models relative to the best ones are displayed in Figure S3-5.

**Table 2.** The identified descriptors and the coefficients in corresponding SISSO models for $BE_H$, $E_b$, and SE.

| property | $d^m$ | descriptor | coefficient |
|---|---|---|---|
| $BE_H$ | $d_1^5$ | (EA*+2F−EC)·DT*·EH*/(EC*+F*) | 0.12653E+00 |
| | $d_2^5$ | $\sqrt[3]{DC}$·H*·DT*·(\|EA*−EH*\|−\|EC−EC*\|) | -0.20440E−02 |
| | $d_3^5$ | \|EH*−L*−\|EH−F*\|\|/(DC$^2$+EC·EC*) | -0.50891E+00 |
| | $d_4^5$ | \|EH−F*−EH*\|−\|EC*−EC−\|DT*−F*\|\| | 0.34705E−01 |
| | $d_5^5$ | L·EC·(EA*+DS*−\|H−EH\|/\|L*−EH*\| | -0.48772E−04 |
| $E_b$ | $d_1^6$ | \|((IP*−L)−\|EC*−DT*\|)/\|EC/DC−L*/IP*\| | -0.87339E−01 |
| | $d_2^6$ | (EA*+DC*+\|DC−DT*\|)/(EA*+EH*+\|L*−F*\|) | -0.19577E−01 |
| | $d_3^6$ | (DC+EH*)·(EC*−F*)·(\|L−EC\|−\|EC−EH\|) | -0.13173E−01 |
| | $d_4^6$ | (DT*−EH)·DC·(H/EC+EA*/L*)/EC* | -0.19172E−01 |
| | $d_5^6$ | $e^{EC}$·EH·DS*/((L*−DS*)+\|H*−EC*\|) | 0.33549E−01 |
| | $d_6^6$ | DC$^2$·(EC*−F*)/(DT*−F*−EA+EC) | -0.14362E−02 |
| SE | $d_1^6$ | (EC+IP+\|F*−DT*\|)/(IP*/R+H*/dd*) | -0.82665E+00 |
| | $d_2^6$ | \|DC−EB*\|·(L−DC−EC)/EB$^2$ | 0.30742E+00 |
| | $d_3^6$ | \|\|EC*−L*\|+\|DC−DS*\|−\|DC−F*\|−\|EC−F*\|\| | 0.11317E+00 |
| | $d_4^6$ | \|H−IP−L+IP*\|/((DC/EC)+(EC/H)) | 0.17455E+00 |
| | $d_5^6$ | (F*−EC)·(L*−DT*−IP)/(F*−EB*) | -0.51761E−02 |
| | $d_6^6$ | EC*·DC·(EB*−L)·(L+L*−EC−DS*) | -0.80032E−03 |

From the Table 2 one can see that the *d*-band center features DC, DC*, DT, DT*, DS, and DS* appear in every dimension of the descriptors for $BE_H$ and $E_b$, consistent with the well-established importance of *d*-band center for adsorption at transition-metal surfaces.[21, 23, 36, 44] The cohesive energies of guest (EC) and host (EC*) bulk metals are selected in each dimension of the descriptor for SE. This is due to the fact that the segregation is driven by the imbalance of binding energy between host and guest-host atoms. Interestingly, most of the descriptor components include only simple mathematical operators (+, −, ·, /, ||), indicating that the primary features already capture most of the complexity of the target properties.

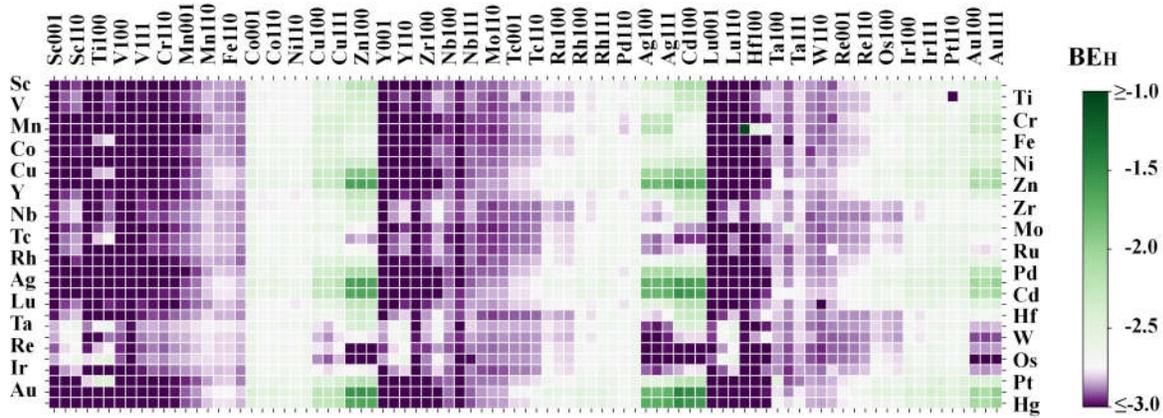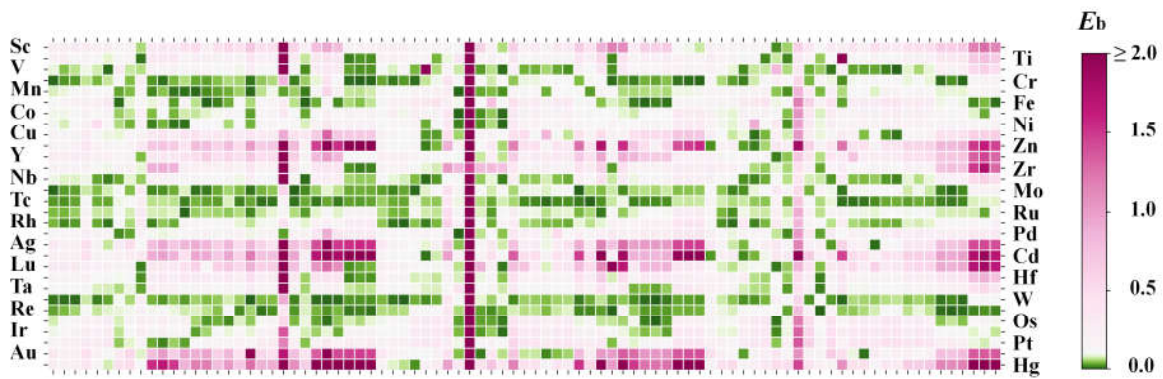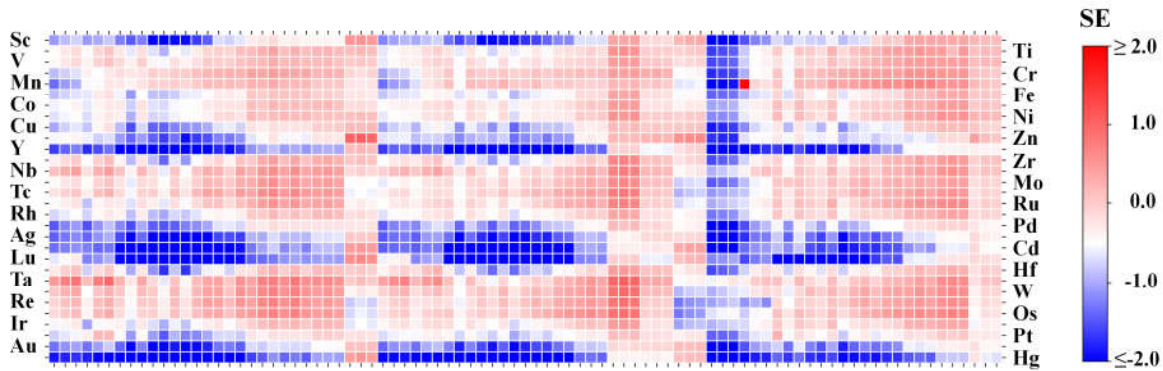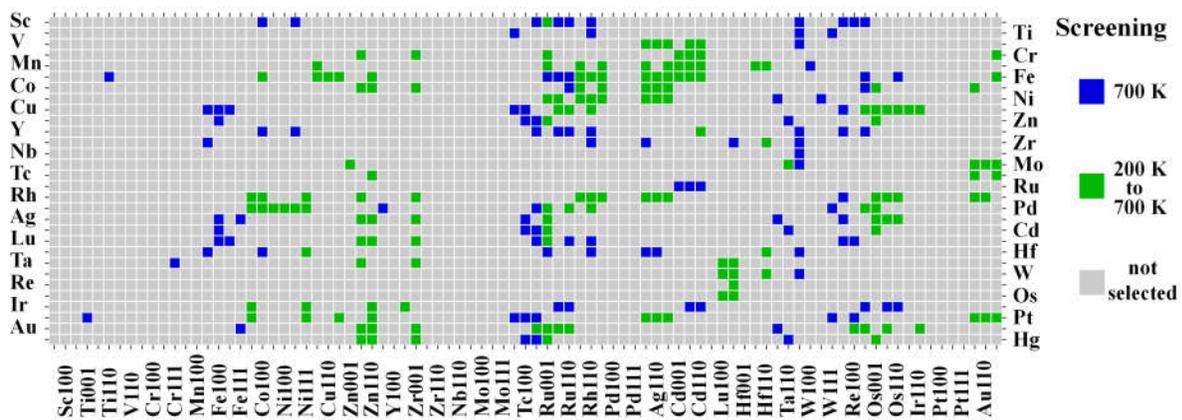

**Figure 3.** High-throughput screening of SAACs for (a) $BE_H$, (b) $E_b$, and (c) SE. The promising candidates at different temperatures $T$ are highlighted in (d). Vertical axis displays the guest atom type, and the horizontal axis displays the host metal surfaces with different surface cuts.

We employ the identified computationally cheap SISSO models to perform high-throughput screening of SAACs to find the best candidates for the hydrogenation reactions. The results for $BE_H$, $E_b$, and SE of the flat surfaces are displayed in Figure 4 a-c (see Figure S6 for the results for the stepped surfaces, the values of $BE_H$, $E_b$, and SE for all the SAACs are given in SI.).

The choice of the screening criteria for the three properties $BE_H$, $E_b$, and SE, which are related to the activity and stability of SAACs, plays the central role in the screening processes and determines the candidates to be chosen. Previous work demonstrates that for the high performance in hydrogenation reactions, SAACs should exhibit weaker binding of H and lower $H_2$ dissociation energy barrier simultaneously.[2] However, different criteria are applicable for different reaction conditions. For example, at low temperatures SAACs can maintain their stability for a longer time. At higher temperatures H atoms will desorb from the surfaces and larger energy barriers can be overcome, resulting in a requirement for stronger binding and higher upper limit of the dissociation barrier $E_b$. Keeping this variability in mind, we consider temperature and pressure-dependent selection criteria (see methods part for details on the selection criteria). We have screened more than five thousand SAAC candidates (including about the same number of flat and stepped surfaces; the values of the primary features for all the candidates can be found in the SI) at both low temperature (200 K) and high temperature (700 K) at partial $H_2$ pressure $p$ = 1 atm. We find 160 flat-surface SAACs (Figure 3d, in green) and 134 stepped-surface SAACs (Figure S6d, in green) that are both active and stable at a low temperature (200 K). At a higher temperature (700 K), 102 flat-surface SAACs (Figure 3d, in blue and green) and 136 stepped-surface SAACs (Figure S6d, in blue and green) are classified as promising SAACs for hydrogenation reactions. Moreover, we have identified the SAACs which are promising in a wide range of temperatures (green squares in Figure 3d for flat surfaces and Figure S6d for stepped surfaces).

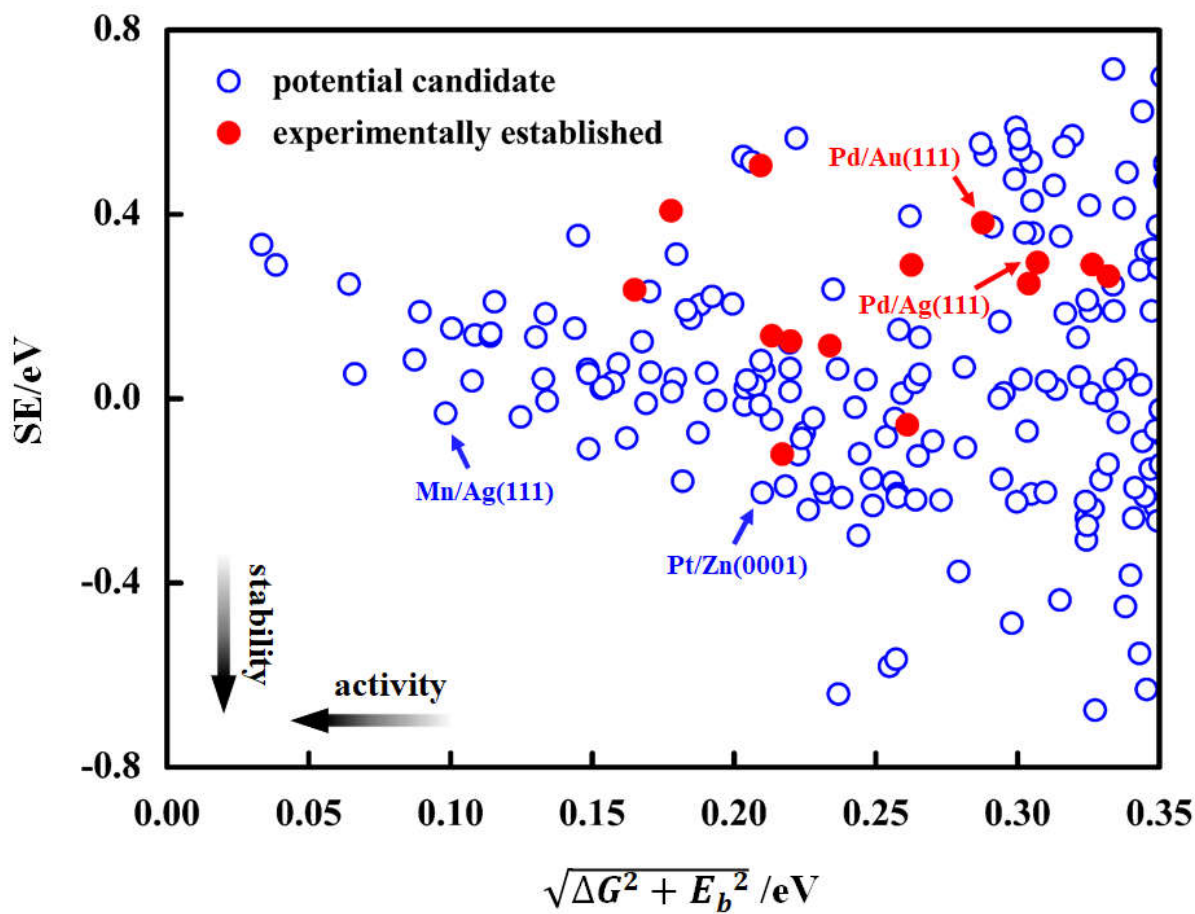

**Figure 4.** Stability vs. activity map for flat SAACs surfaces at $T = 298\,K$ and $p = 1$ atm. The SE on y-axis represents stability and activity parameter $\sqrt{\Delta G^2 + E_b^2}$ is shown on x-axis. Experimentally established SAACs are denoted with red solid spheres and the blue open circles represent new predicted candidates.

Note that, without the stability selection criterion based on SE, all experimentally established SAACs (Pd/Cu, Pt/Cu, Pd/Ag, Pd/Au, Pt/Au, Pt/Ni, Au/Ru, Ni/Zn) are predicted to be good catalysts in the temperature range of 200 K < T < 700 K, which are further confirmed by DFT calculations. However, some of these systems (Pd/Ag and Pd/Au) are experimentally shown to have low stability.[12, 16] Thus, inclusion of the stability-related property SE is of immense importance for a reliable prediction of catalytic performance, as is confirmed by our results. We define an activity (or efficiency) indicator involving both the free energy of H adsorption ($\Delta G$) and the energy barrier ($E_b$) as $\sqrt{\Delta G^2 + E_b^2}$ to construct an activity-stability map. As shown in Figure 4, some of the new discovered candidates (bottom-left corner of activity-stability map) are predicted to have both higher stability and efficiency than the reported ones, making them perfectly optimized for practical applications (see Figure S7 for the results for the stepped surfaces). Considering stability, activity, and abundance, two discovered best candidates Mn/Ag(111) and Pt/Zn(0001) are highlighted in Figure 4.

In summary, by combining first-principles calculations and the data-analytics approach SISSO, we have identified accurate and reliable models for the description of the hydrogen binding energy, dissociation energy, and guest-atom segregation energy for SAACs, which allow us to make fast yet reliable prediction of the catalytic performance of hundreds of thousands SAACs in hydrogenation reactions. The model correctly evaluates performance of experimentally tested SAACs. By scanning more than five thousand SAACs with our model, we have identified over two hundred new SAACs with both improved stability and performance compared to the existing ones. Our approach can be easily adapted to designing new functional materials for various applications.

**Methods**

All first-principles calculations are performed with the revised Perdew-Burke-Ernzerhof (RPBE) functional[53] as implemented in the all-electron full-potential electronic-structure code FHI-aims.[54] The choice of functional is validated based on a comparison of calculated $H_2$ adsorption energies to the available experimental results[55] (see Table S1). Nevertheless, it is expected that, because of the large set of systems inspected and the small variations introduced by the functional choice, the main trends will hold even when using another functional (see Supporting Information (SI) for more details on the computational setup). The climbing-image nudged elastic band (CI-NEB) algorithm is employed to identify the transition state structures.[56]

$BE_H$ are calculated using equation (1), where $E_{H/support}$ is the energy of the total H/support system, $E_{support}$ is the energy of the metal alloy support, and $E_H$ is the energy of an isolated H atom.

$$BE_H = E_{H/support} - E_{support} - E_H \qquad (1)$$

The surface segregation energy in the dilute limit, SE, is defined as the energy difference of moving the single impurity from the bulk to the surface when surface H adatom is present (the H is put at the most stable adsorption site for each system). In this work, it is calculated using equation (2), where $E_{top-layer}$ and $E_{nth-layer}$ correspond to the total RPBE energy of the slab with the impurity in the top and $n$th surface layer, respectively. The value of $n$ is chosen so that the energy difference between $E_{nth-layer}$ and $E_{(n-1)th-layer}$ is less than 0.05 eV.

$$SE = E_{surface} - E_{nth-layer} \qquad (2)$$

Using first-principles inputs as training data, we have employed SISSO to single out a simple and physically intuitive descriptor from a huge number of potential candidates. In practice, a huge pool of more than ten billion candidate descriptors is first constructed iteratively by combining user-defined primary features with a set of mathematical operators. The number of times the operators are applied determines the complexity of the resulting descriptors. We consider up to three levels of complexity (feature spaces) $\Phi_1$, $\Phi_2$, and $\Phi_3$. Note that a given feature space $\Phi_n$ also contains all of the lower rung (i.e. $n$-1) feature spaces. Subsequently, the desired low-dimensional representation is obtained from this pool.[49] The details of the feature space ($\Phi_n$) construction and the descriptor identification processes can be found in the SI. The proper selection of primary features is crucial for the performance of SISSO-identified descriptors. Inspired by previous studies,[31, 38] we consider three classes of primary features (see

Table 1) related to the metal atom, bulk, and surface. The more detailed description and values of all the primary features are given in the SI.

The selection of the promising candidates at various temperatures and hydrogen partial pressures is performed based on *ab initio* atomistic thermodynamics.[57] H adsorption/desorption on SAAC surfaces as a function of temperature and $H_2$ partial pressure $(T, p)$ is characterized by the free energy of adsorption $\Delta G$:

$$\Delta G = E_{H/support} - E_{support} - \mu_H(T, p)$$

with the chemical potential of hydrogen $\mu_H = \frac{1}{2}\mu_{H_2}$ obtained from:

$$\mu_H = \frac{1}{2}\left(E_{H_2} + \Delta\mu_{H_2}(T, p)\right)$$

where $\Delta\mu_{H_2}(T, p) = \mu_{H_2}(T, p^0) - \mu_{H_2}(T^0, p^0) + k_B T \ln(\frac{p}{p^0})$.

Here $T^0 = 298\ K$ and $p^0 = 1$ atm. The first two terms are taken from JANAF thermochemical tables.[58] In the following, we set $p = 1$ atm.

According to Sabatier principle the optimum heterogeneous catalyst should bind the reactants strong enough to allow for adsorption, but also weak enough to allow for the consecutive desorption.[25] In this work, a $BE_H$ range is defined by the conditions:

$$|BE_H - \frac{1}{2}(E_{H_2} - 2E_H) - \frac{1}{2}\Delta\mu_{H_2}(T)| < 0.3\ \text{eV}$$

where $E_{H_2} - 2E_H$ is the hydrogen binding energy of the hydrogen molecule. The experimental value of -4.52 eV[59] was used in this work.

The above conditions correspond to the free-energy bounds:

$$|\Delta G| < 0.3\ \text{eV},$$

Conditions on energy barrier ($E_b$) are defined by considering Arrhenius-type behaviour of the reaction rate on $E_b$ and $T$. Assuming that acceptable barriers are below 0.3 eV for $T^0 = 298$ K, we estimate acceptable barrier at any temperature as $E_b < \frac{0.3T}{T^0}$ eV.

Similarly the bounds for SE are determined by imposing a minimum 10% ratio for top-layer to subsurface-layers dopant concentration by assuming an Arrhenius-type relation with SE interpreted as activation energy:

$$SE < k_B T \ln(10)$$

**Acknowledgments**

S.V.L. is supported by Skolkovo Foundation Grant.

Y.G. is supported by the National Natural Science Foundation of China (11604357, 11574340).

R.O. is supported by the National Key Research and Development Program of China (2018YFB0704400) and the Program of Shanghai Youth Oriental Scholars.

## Author contributions

S.V.L. created the idea and conceived the work. S.V.L. and Y.G. designed and supervised the project. Z.-K.H. performed all the calculations. Z.-K.H., D.S., O.R., Y.G., and S.V.L. co-wrote the manuscript. All authors contributed to the analysis and interpretation of the results. All the authors commented on the manuscript and have given approval to the final version of the manuscript.

## Competing interests

The authors declare no competing financial interests.

## Additional information

Supplementary Information is available for this paper at http://www.nature.com/nature.

Correspondence and requests for materials should be addressed to S.V.L. or Y.G.

# Single-Atom Alloy Catalysts Designed by First-Principles Calculations and Artificial Intelligence

Zhong-Kang Han,[1,4] Debalaya Sarker,[1,4] Runhai Ouyang,[2,4] Yi Gao,[3]* Sergey V. Levchenko[1]*

[1] Center for Energy Science and Technology, Skolkovo Institute of Science and Technology, Skolkovo Innovation Center, Moscow, 143026, Russia
[2] Materials Genome Institute, Shanghai University, 333 Nanchen Road, Shanghai, 200444, P.R. China
[3] Shanghai Advanced Research Institute, Chinese Academy of Sciences, Shanghai, 201210, P. R. China
[4] These authors contributed equally.

Correspondence Email: S.Levchenko@skoltech.ru; gaoyi@sinap.ac.cn

## S1. Computational details

All *ab initio* calculations were performed with the revised Perdew-Burke-Ernzerhof (RPBE) functional[1] as implemented in the all-electron full-potential electronic-structure code FHI-aims[2] using density-functional theory (DFT) and numerical atom-centered basis functions. The numerical settings are so chosen for the present study that a convergence better than $10^{-3}$ eV/atom in energy differences is achieved. The standard 'tight' settings (grids and basis functions) were employed which deliver the adsorption energies with basis-set superposition errors below 0.07 eV per adsorbed molecule. The choice of functional is validated based on a comparison of calculated $H_2$ adsorption energies to the available experimental results[3] (Table S1). Spin-polarization effects are tested for and included where appropriate. Slabs of at least nine metal layers were considered with the two to four bottom layers fixed, based on the convergence of $BE_H$ and SE (within 0.05 eV) with respect to the thickness and supercell size of the slab. The lattice vector along the direction parallel to the vacuum gap was 50 Å. All atoms in the systems except for the fixed bottom have been allowed to relax until the maximum remaining force fell below $10^{-2}$ eV/Å. The climbing-image nudged elastic band (CI-NEB) algorithm is employed to identify the transition state structures.[4] H atom is placed at different non-equivalent high-symmetry sites close to the guest atom (Figure S1), and the $BE_H$ for the most favorable site is included in the data set. The host metal surfaces considered in this work to construct the training data set are Cu(100), Cu(310), Zn(0001), Cr(110), Pd(111), Pd(211), Pt(111), Rh(111), Ru(0001), Cd(0001), Ag(100), Ag(110), Ti(0001), Nb(210), and Ta(210) with more than three hundred points for each considered properties. All the DFT calculated $BE_H$, SE, and $E_b$ can be found in the file "data.txt".

## S2. Additional SISSO computational details

For constructing the $\Phi_1$, $\Phi_2$, and $\Phi_3$ feature spaces we made use of the set of algebraic/functional operators given in eq. 1.

$$\hat{H}^{(m)} \equiv \{+, -, \cdot, /, \log, \exp, \exp-, ^{-1}, ^2, ^3, \sqrt{}, \sqrt[3]{}, |-|\}, \qquad (1)$$

The superscript *m* indicates that when applying $\hat{H}^{(m)}$ to primary features $\varphi_1$ and $\varphi_2$ a dimensional analysis is performed, which ensures that only physically meaningful combinations are retained

(e.g. only primary features with the same unit are added or subtracted). All primary features included in this study were obtained either from the literature (see Table S2), or from DFT calculations (see Table S3). The values of the primary features for the training data sets can be found in the file "data.txt" and the values of the primary features for all the high-throughput screening SAAC candidates can be found in the file "high-throughput-data.txt".

The sparsifying $\ell_0$ constraint is applied to a smaller feature subspace selected by a screening procedure (sure independence screening (SIS)), where the size of the subspace is equal to a user-defined SIS value times the dimension of the descriptor. The SIS value is not an ordinary hyperparameter and its optimization through a validation data set is not straightforward. Ideally, one would want to search the entire feature space for the optimal descriptor. However, this is not computationally tractable since the computational cost of the sparsifying $\ell_0$ constraint grows exponentially with the size of the searched feature space. Instead, the SIS value should be chosen as large as computationally possible. The reasonable SIS values were chosen based on the convergence of the training error.

To confirm the reliability of the SISSO model optimization approaches, the data were initially divided into training and test sets. The Pd(211) and Pt(111) based systems are used as test set while all the other data were contained in the training set. As mentioned in the main text, 10-fold cross validation (CV10) method was used to determine the dimensionality of the descriptor. First, the best descriptors were selected by SISSO based on only the training data. The RMSR and CV10 errors on the training set for the descriptors of $BE_H$, $E_b$, and SE are displayed in Figure S2a. Second, the predictive power of the SISSO selected descriptors was tested using the test set. We display in Figure S2b the distribution of errors on the training set and test set for the descriptors of $BE_H$ (green box), $E_b$ (blue box), and SE (cyan box). To check the predictive power of SISSO selected descriptors on different types of surfaces, we divide the test set into two groups: one with only flat surface and the other one with only stepped surface. In the first group, we considered a new transition metal of Pt which are not contained in the training set (Cu, Zn, Cr, Pd, Rh, Ru, Cd, Ag, Ti, Nb, and Ta), while in the second group, we considered a new surface cut of fcc(211) which is also not included in the training set. The root-mean-square errors, RMSEs, (maximum absolute errors, MAEs) of the SISSO selected descriptors for the test flat surface set are found to be $BE_H$: 0.10 eV (0.43 eV), $E_b$: 0.12 eV (0.62 eV), and SE: 0.22 eV (0.78 eV), while for the test stepped surface set are found to be $BE_H$: 0.11 eV (0.54 eV), $E_b$: 0.14 eV (0.71 eV), and SE: 0.24 eV (0.87 eV). The moderate errors of the RMSEs for both test flat surface set and test stepped surface set showed that the transferability of the descriptors is good. The little larger RMSEs and MAEs for test stepped surface set compared to that for test flat surface set can be rationalized by the fact that in the training set, more flat surfaces are considered compared to stepped surfaces. We would expect the errors for the test set to further decrease when more data are included in the training set. Thus, after confirming the reliability of the of the SISSO model optimization approaches, finally we included all the data into the training set for the SISSO selection of the best descriptors. As mentioned in the main text, the obtained optimal descriptor dimensionalities for $BE_H$, $E_b$, and SE of the SAACs are 5, 6, and 6, respectively. The same optimal descriptor dimensionalities for $BE_H$, $E_b$, and SE were found when the Pt(111) and Pd(211) based systems are not included, further confirming the reliability the used SISSO model optimization approaches. The error distributions for all the lower dimensional models are displayed in Figure S3-5. The RMSE and MAE of the models are also shown. The identified lower-dimensional descriptors and the coefficients in corresponding SISSO models for $BE_H$, $E_b$, and SE can be found in the file "descriptor.txt". The top five largest deviations between calculated and predicted $BE_H$, $E_b$, and SE are collected in Table S4. As can be seen, the deviation for SE is larger than that of $BE_H$ and $E_b$. However, we found even for SE our model's precision is higher than 95% (Table S5).

**Table S1.** The experimental and theoretical adsorption energies (in eV) of $H_2$ at different transition metal surfaces.

| method | Pt(111) | Ru(0001) | Pd(111) |
|---|---|---|---|
| RPBE | -0.56 | -1.06 | -1.01 |
| PBE | -0.32 | -0.78 | -0.70 |
| Experiments[3] | -0.75 | -1.20 | -0.91 |

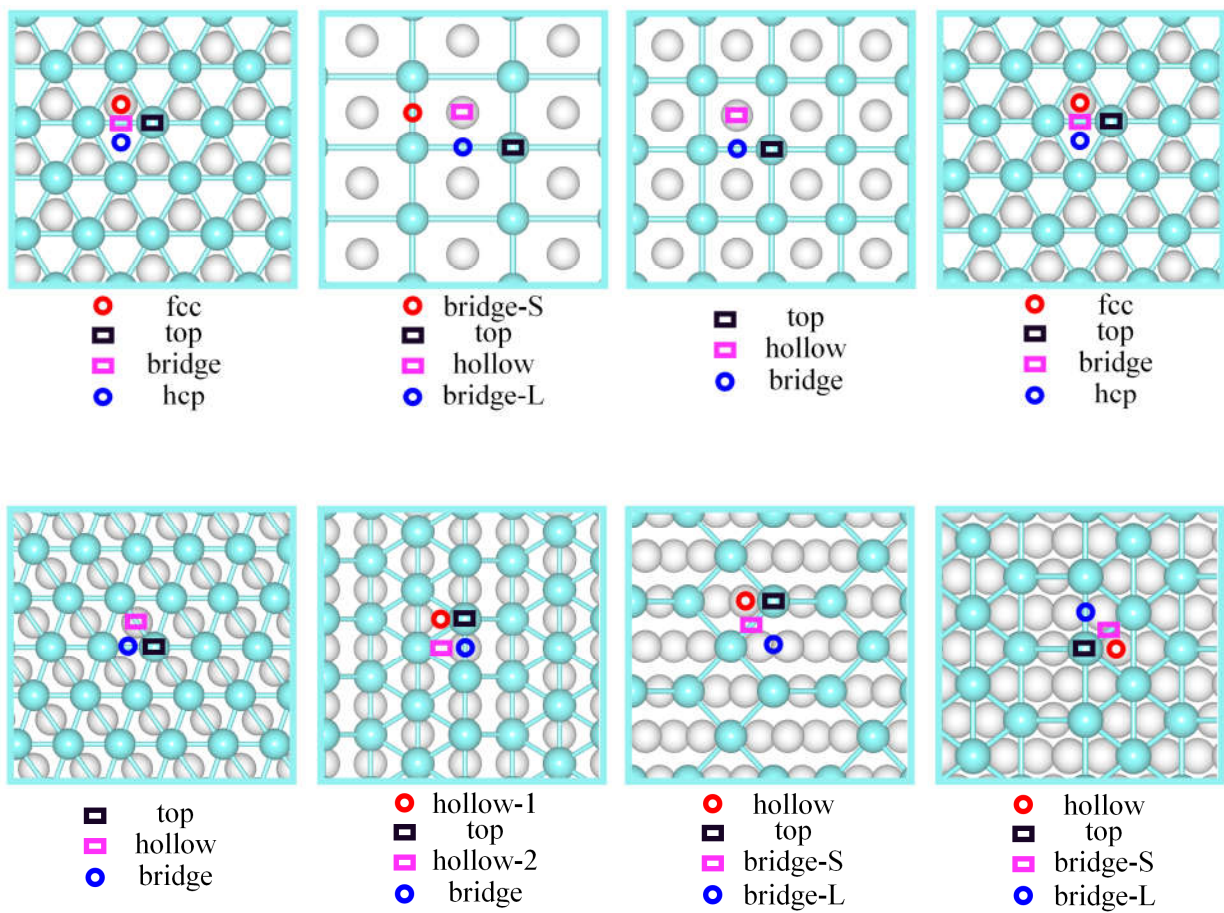

**Figure S1.** The considered hydrogen adsorption sites on the fcc(111) (a), fcc(110) (b), fcc(100) (c), hcp(0001) (d), bcc(110) e, fcc(211) (f), fcc(310) (g), bcc(210) (h) of pure transition metal

surfaces. The atom below the top site for each surface facet is either the host atom (pure transition metal surfaces) or the single guest atom (single atom alloy metal surfaces).

**Table S2.** Primary features obtained from the literature.[5] Electron affinity (EA in eV), ionization potential (IP in eV), and covalent radius (R) of the metal atom.

| system | class | name | abbreviation |
|---|---|---|---|
| host | atomic | Electron affinity | EA* |
| | | Ionization potential | IP* |
| | | Atomic radius | R* |
| guest atom | atomic | Electron affinity | EA |
| | | Ionization potential | IP |
| | | Atomic radius | R |

**Table S3.** Primary features obtained from DFT-RPBE calculations (spin-polarization effects are tested for and included where appropriate). Energy of the highest-occupied Kohn-Sham level (H in eV), energy of the lowest-unoccupied Kohn-Sham level (L in eV), binding energy of H with isolated metal atom (EH in eV as calculated by equation (2)), binding energy of metal dimers (EB in eV as calculated by equation (3)), binding distance of H with isolated metal atom (dH in Å), and binding distance of metal dimers of the metal atom; cohesive energy (EC in eV as calculated by equation (4)) and $d$-band center (DC in eV) of the bulk metal; $d$-band center of the top surface layer (DT in eV), $d$-band center of the subsurface layer (DS in eV), and the slab Fermi level (F in eV) of the metal surface.

| system | class | name | abbreviation |
|---|---|---|---|
| host | atomic | Energy of the highest-occupied Kohn-Sham level | H* |
| | | Energy of the lowest-unoccupied Kohn-Sham level | L* |
| | | Binding energy of H with single host metal atom | EH* |
| | | Binding energy of host metal dimers | EB* |
| | | Binding distance of H with single host metal atom | dH* |
| | | Binding distance of host metal dimer | dd* |
| | bulk | Cohesive energy | EC* |
| | | $d$-band center | DC* |
| | surface[#] | $d$-band center of the top surface layer | DT* |
| | | $d$-band center of the subsurface layer | DS* |
| | | Slab Fermi level | F* |
| guest atom | atomic | Energy of the highest-occupied Kohn-Sham level | H |
| | | Energy of the lowest-unoccupied Kohn-Sham level | L |
| | | Binding energy of H with single guest metal atom | EH |
| | | Binding energy of guest metal dimers | EB |
| | | Binding distance of H with single guest metal atom | dH |

|  | Binding distance of guest metal dimers | dd |
| bulk | Cohesive energy | EC |
|  | d-band center | DC |

[#]the surface based primary features were calculated by using the unit cell consisting of one atom per atomic layer.

$EH = E_{H\text{-metal}} - E_{metal} - E_H$          (2)

where $E_{H\text{-metal}}$ is the energy of the total H-metal system, $E_{metal}$ is the energy of the isolated metal atom, and $E_H$ is the energy of the isolated H atom.

$EB = E_{metal\text{-metal}} - 2E_{metal}$          (3)

where $E_{metal\text{-metal}}$ is the energy of the total metal-metal dimer system, $E_{metal}$ is the energy of the isolated metal atom.

$EC = E_{bulk}/n - E_{metal}$          (4)

where $E_{bulk}$ is the energy of the bulk metal system, $E_{metal}$ is the energy of the isolated metal atom, and $n$ is the number of metal atoms in the bulk unit cell.

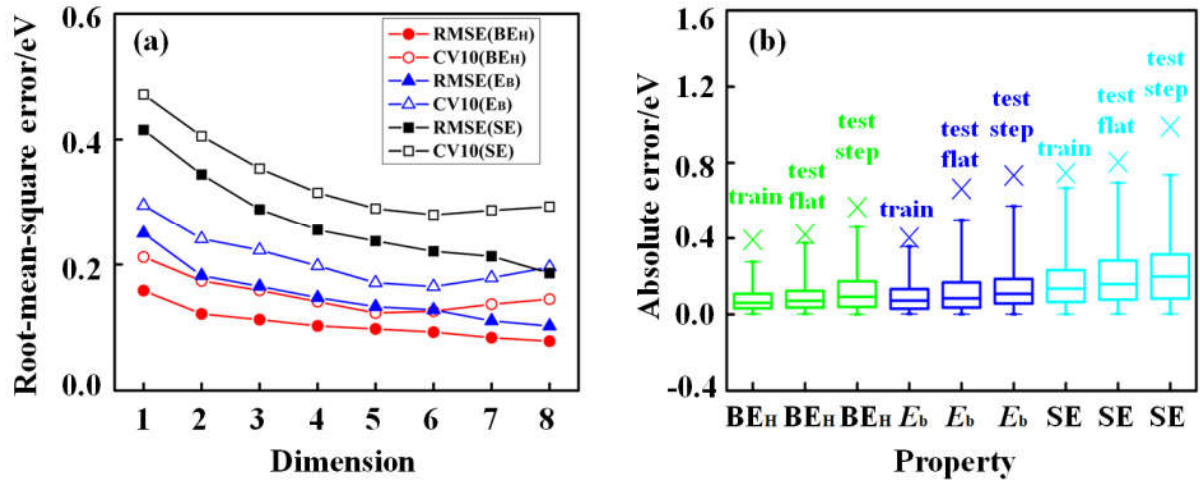

**Figure S2.** (a) RMSE and the averaged RMSE of the 10 fold cross-validation. (b) Box plots of the absolute errors for the training set and test set for the SISSO selected best models of $BE_H$ (green), $E_b$ (blue), and SE (cyan). The test set is divided into two parts: one part contains only flat surfaces and the other one contains only stepped surfaces. The upper and lower limits of the rectangles represent the 75th and 25th percentiles of the distribution, the internal horizontal lines mark the median (50th percentile), and the upper and lower limits of the error bars indicate the 99th and 1st percentiles. The crosses depict the maximum absolute errors.

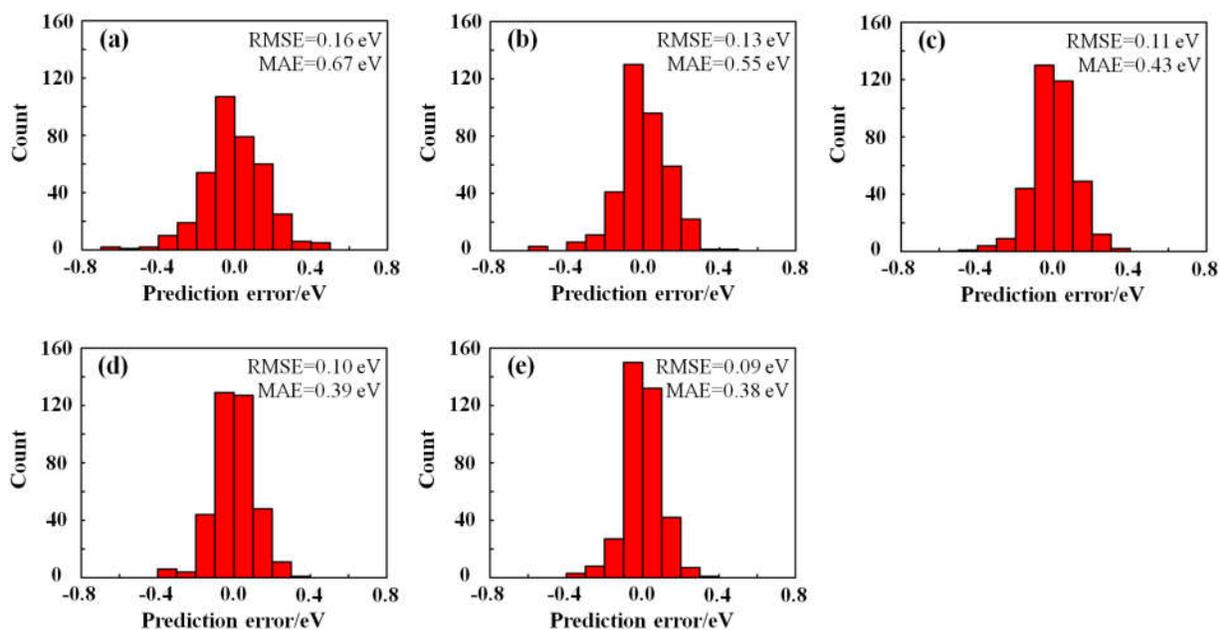

**Figure S3.** The error distributions for all the lower-dimensional models of $BE_H$: (a) 1D, (b) 2D, (c) 3D, and (d) 4D, and the SISSO selected best model (e) 5D.

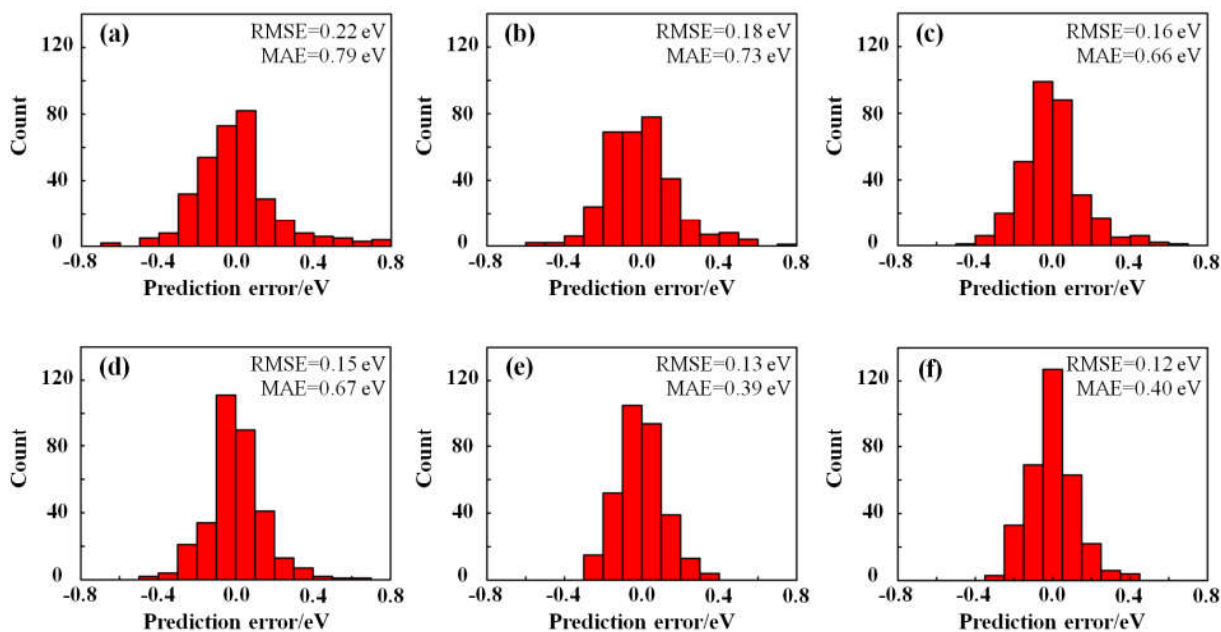

**Figure S4.** The error distributions for all the lower dimensional models of $E_b$: (a) 1D, (b) 2D, (c) 3D, (d) 4D, (e) 5D and the SISSO selected best model (f) 6D.

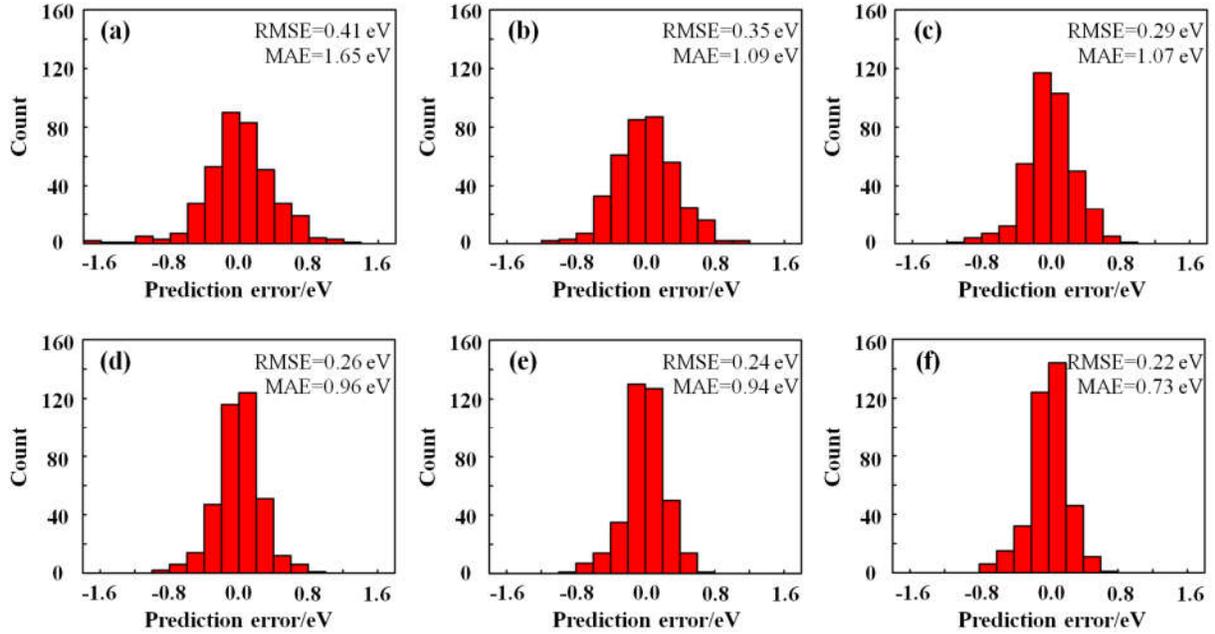

**Figure S5.** The error distributions for all the lower dimensional models of SE: (a) 1D, (b) 2D, (c) 3D, (d) 4D, (e) 5D and the SISSO selected best model (f) 6D.

**Table S4.** The top five largest deviations between calculated and predicted $BE_H$ (in eV), $E_b$ (in eV), and SE (in eV).

| property | system | calculated | predicted | deviation |
|---|---|---|---|---|
| $BE_H$ | Co/Ag(100) | -2.98 | -2.60 | 0.38 |
| | Co/Ag(110) | -2.98 | -2.65 | 0.33 |
| | Ni/Ti(0001) | -2.95 | -3.27 | 0.33 |
| | Ta/Ti(0001) | -3.19 | -2.89 | 0.30 |
| | Sc/Zn(0001) | -1.91 | -2.21 | 0.30 |
| $E_b$ | Hf/Pt(111) | 0.78 | 0.38 | 0.40 |
| | Cu/Ag(110) | 0.92 | 0.52 | 0.40 |
| | Ir/Zr(0001) | 0.68 | 0.28 | 0.40 |
| | Pd/Ag(110) | 0.60 | 0.24 | 0.36 |
| | Zr/Cu(100) | 0.71 | 0.38 | 0.33 |
| SE | V/Pd(111) | -0.27 | 0.46 | 0.73 |
| | Hg/Zn(0001) | -0.24 | 0.48 | 0.71 |
| | Os/Zn(0001) | 0.96 | 0.25 | 0.71 |
| | Cd/Zn(0001) | -0.13 | 0.50 | 0.62 |
| | Hg/Cd(0001) | -0.19 | 0.43 | 0.62 |

Table S5. Number of system with the predicted and calculated segregation energy meet the same condition of SE < kTln(10) ($N_{meet}$), the total number of calculated systems ($N_{total}$), and the SE model's precision ($P = N_{meet}/ N_{total}$).

| temperature | $N_{meet}$ | $N_{total}$ | $P$ |
|---|---|---|---|
| 200 K | 345 | 360 | 95.83% |
| 700 K | 346 | 360 | 96.11% |

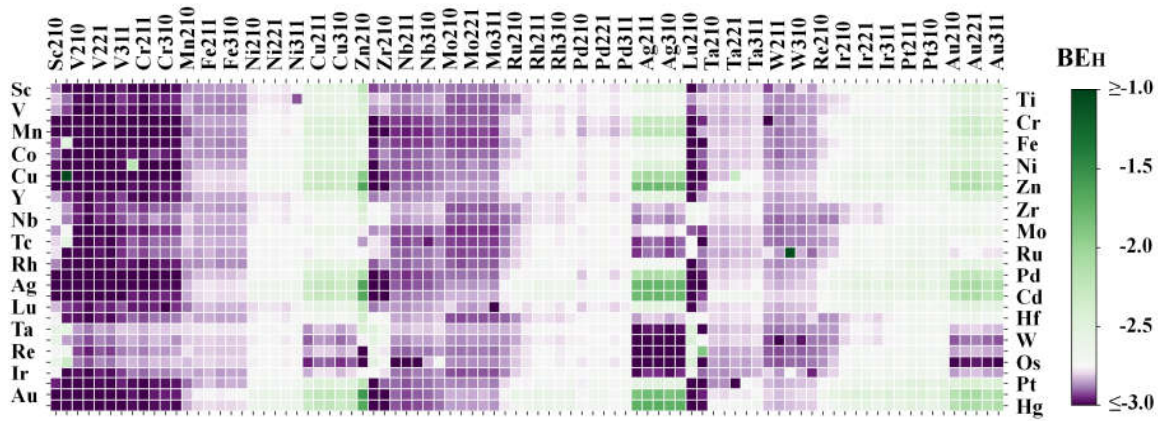
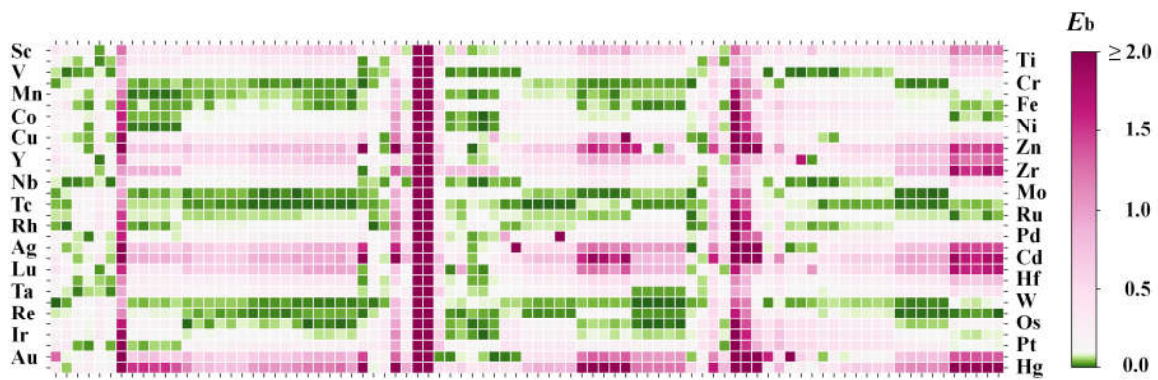
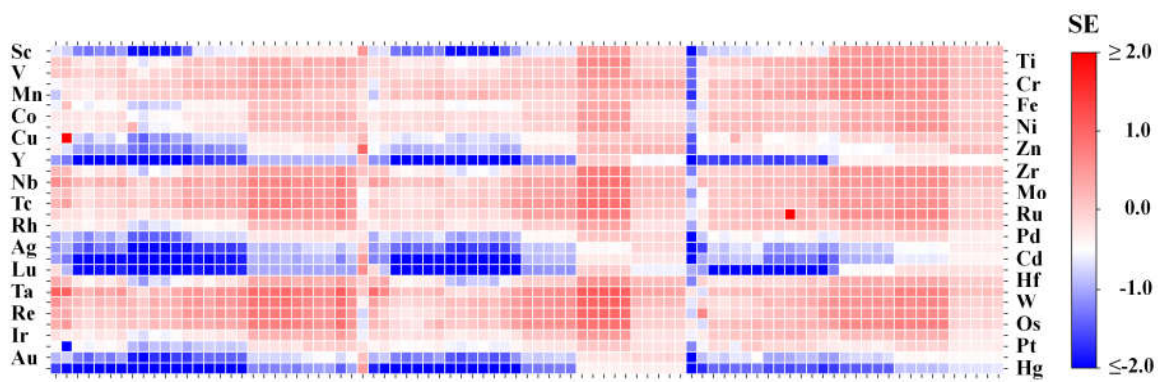
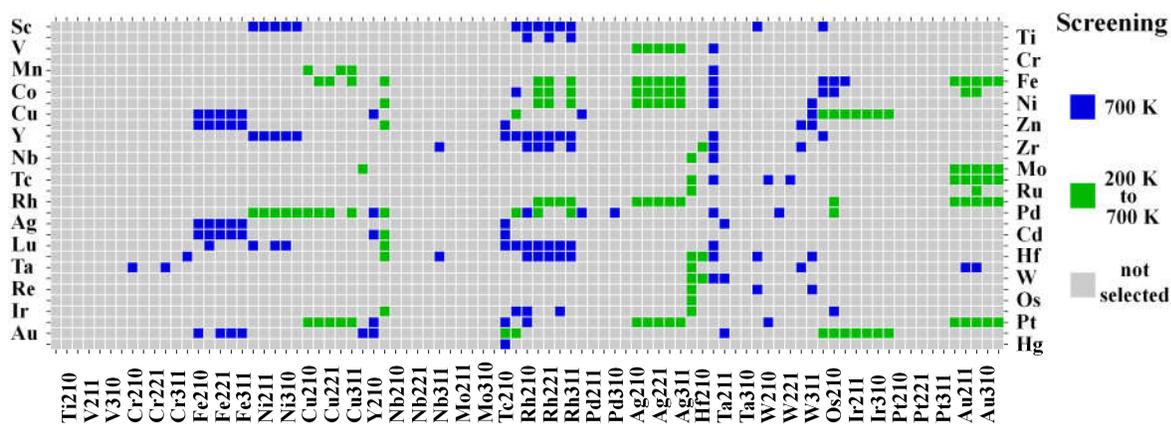

**Figure S6.** High-throughput screening of SAACs for (a) $BE_H$, (b) $E_b$, and (c) SE. The screened candidates are highlighted in (d). Vertical axis displays the guest atom type, and the vertical horizontal axis displays the host metal surfaces with different stepped surface cuts.

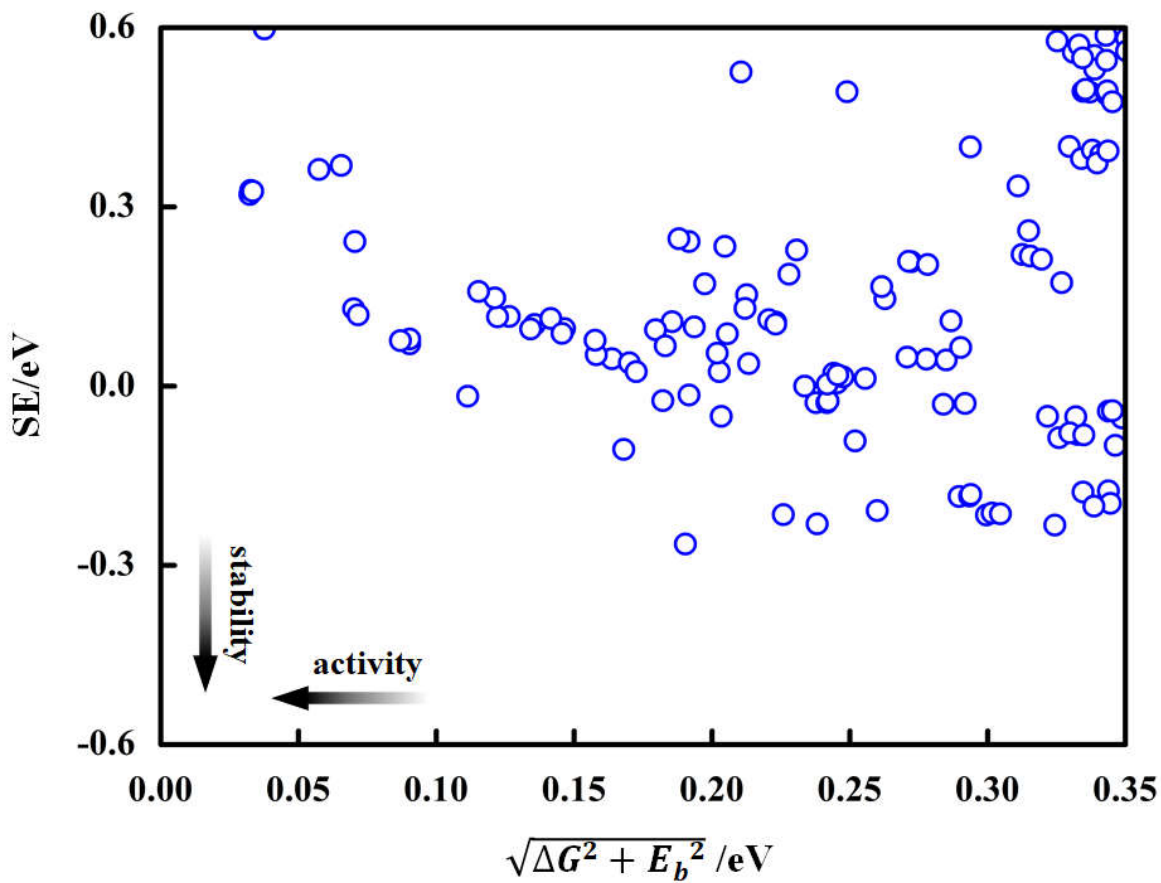

**Figure S7.** Stability vs. activity map for step SAACs surfaces at $T = 298\ K$ and $p = 1$ atm. The SE on y-axis represents stability and activity parameter $\sqrt{\Delta G^2 + E_b^2}$ is shown on x-axis.